\documentclass{esannV2}
\usepackage[dvips]{graphicx}
\usepackage[latin1]{inputenc}
\usepackage{amssymb,amsmath,array}
\usepackage{epsfig}

\usepackage{dsfont}
\usepackage{amssymb}
\usepackage{amsmath}
\usepackage{graphics}
\usepackage{graphicx}
\usepackage{fancyhdr}
\usepackage[english]{babel}
\usepackage[usenames, dvipsnames]{color}
\usepackage[perpage]{footmisc}

\usepackage{ifthen}
\usepackage{ifpdf}

\usepackage{algorithmic}
\usepackage[ruled]{algorithm}
\usepackage{enumerate}
\usepackage{theorem}
\usepackage[subnum]{cases}
\usepackage{multirow}
\usepackage{color}

\newcommand{\bh}{\mathbf{h}}

\newcommand{\bt}{\mathbf{t}}

\newcommand{\bT}{\mathbf{T}}

\newcommand{\bw}{\mathbf{w}}

\newcommand{\bx}{\mathbf{x}}

\newcommand{\bz}{\mathbf{z}}

\newcommand{\bsw}{\boldsymbol{w}}

\newcommand{\cL}{\mathcal{L}}

\newcommand{\bsbeta}{\boldsymbol{\beta}}

\newcommand{\bstheta}{\boldsymbol{\theta}}

\newcommand{\bsPsi}{\boldsymbol{\Psi}}

\newcommand{\Identity}{\textbf{I}}

\newcommand{\E}{\mathbb{E}}

\newcommand{\N}{\mathcal{N}}

\graphicspath{{Figures/}}

\begin{document}
\title{Functional Mixture Discriminant Analysis with hidden process regression for curve classification}

\author{Faicel Chamroukhi$^1$, Herv{\'e} Glotin$^{1,2}$, C{\'e}line Rabouy$^1$ 
%
\thanks{{\small Work supported by the  ANR  project ``Cognilego" 2010-CORD-013.}}
%
\vspace{.3cm}\\
%
1- Information Sciences and Systems Laboratory (LSIS), UMR CNRS  7296\\
University of the South Toulon-Var, 83957 La Garde Cedex, France
\\2- Institut Universitaire de France, iuf.amue.fr
}

\maketitle

\begin{abstract}
We present a new mixture model-based discriminant analysis approach for functional data using a specific hidden process regression model. The approach allows for fitting flexible curve-models to each class of complex-shaped curves presenting regime changes. The model parameters are learned by maximizing the observed-data log-likelihood for each class by using a dedicated expectation-maximization (EM) algorithm. Comparisons on simulated data with alternative approaches show that the proposed approach provides better results. 
\end{abstract}

\section{Introduction}

Most statistical analyses involve vectorial data when the observations are presented by finite dimensional vectors. However, in many application domains, the observations are functions or curves. The general paradigm for analyzing such data is known as `Functional Data Analysis' (FDA) \cite{ramsayandsilvermanFDA2005}. The goals of FDA 
include data representation for further analysis,  data visualization, exploratory analysis by performing unsupervised approaches, classification, etc. 
In this study, we focus on supervised functional data classification (discrimination) where the observations are temporal curves presenting regime changes over time. The problem of curve classification, as in the case of multidimensional vectorial data classification, can be addressed by learning generative or discriminant classifiers. Among the discriminative approaches, one can cite 
the one based on support vector machines \cite{RossiAndVillaSVM_functional_data_Neurocomputing_2006}. The generative approaches are essentially based on regression analysis, including polynomial or
 spline regression 
\cite{chamroukhi_PhD_2010, Gui_FMDA, garetjamesANDtrevorhastieJRSS2001}, or also generative polynomial piecewise regression as in \cite{chamroukhi_PhD_2010, chamroukhi_et_al_neurocomputing2010}. 
%
%
The generative models aim at understanding the process generating such data to handle both the problem of heterogeneity between curves and the process governing the regime changes, in order to fit  flexible models that provide better classification results.
In this paper, we propose a new generative approach for modeling classes of complex-shaped curves where each class is composed of unknown sub-classes. In addition, the model is particularly dedicated to address the problem when each homogeneous sub-class presents regime changes over time. 
We extend the functional discriminant analysis approach presented in \cite{chamroukhi_et_al_neurocomputing2010}, which relates modeling each class of curves presenting regime changes with a single mean curve, to a mixture formulation that leads to a functional mixture-model based discriminant analysis. More specifically, this approach reformulates the entirely unsupervised mixture of regression models with hidden logistic processes (MixRHLP) \cite{chamroukhi_PhD_2010, chamroukhi_adac_2011} into a supervised functional data classification framework. The resulting discrimination approach is a model-based functional discriminant analysis in which each class  of curves is modeled by a MixRHLP model. 
In the next section we give a brief background on discriminant analysis approaches for functional data and then we present the proposed mixture model-based functional discriminant analysis with hidden process regression. 
%
Let us denote by $((\bx_1,y_1),\ldots,(\bx_n,y_n))$ a given labeled training set of curves issued from $G$ classes with $y_i \in  \{1,\ldots,G \}$ the class label of the $i$th curve. 
%
 Assume that each curve $\bx_i$ consists of $m$ observations $\bx_i = (x_{i1},\ldots,x_{im})$, regularly observed at the time points $t_1<\ldots<t_m$. 

\vspace*{-.2cm}

\section{Functional Discriminant Analysis}

Functional discriminant analysis extends discriminant analysis approaches for vectorial data to functional data or curves. From a probabilistic point a view, the class conditional density is assumed to be a (parametric) density defined in the functional space rather than in a finite dimensional space of the multidimensional data vectors. Assume we have  a labeled training set of curves and the class parameter vectors $(\bsPsi_1,\ldots,\bsPsi_G)$ where $\bsPsi_g$ is the parameter vector representing the density of class $g$ $(g=1,\ldots,G)$. 
In functional discriminant analysis, a new curve $\bx_i$ is assigned to the class $\hat{y}_i$ using the Bayes rule, that is: 
{\small \begin{equation}
\hat{y}_i=\arg \max_{1\leq g\leq G} \frac{w_g p(\bx_i|y_i=g,\bt;\bsPsi_g)}{\sum_{g'=1}^{G}w_{g'}p(\bx_i|y_i=g',\bt;\bsPsi_{g'})} ,
\label{eq: MAP rule for FDA classification}
\end{equation}}where $w_g = p(y_i=g)$ is the prior probability of class $g$ 
and $p(\bx_i|y_i=g,\bt;\bsPsi_g)$ its density. The class conditional density 
 can be defined as the one of a polynomial regression model, polynomial spline including B-spline 
\cite{garetjamesANDtrevorhastieJRSS2001}, or a generative piecewise regression model with a hidden process 
 \cite{chamroukhi_et_al_neurocomputing2010} when the curves further present regime changes over time. These approaches lead to functional linear (or quadratic) discriminant analysis. 
%
\subsection{Functional Linear Discriminant Analysis}
\label{ssec: FLDA state of the art}
Functional Linear (or Quadratic) Discriminant Analysis \cite{garetjamesANDtrevorhastieJRSS2001} arises when we model each class conditional density of curves 
$p(\bx_i|y_i=g,\bt;\bsPsi_g)$ by a single model. The class conditional density can for example be the one of a polynomial, spline or B-spline regression model with parameters  $\bsPsi_g$, that is:
{\small \begin{eqnarray}
p(\bx_i|y_i=g,\bt;\bsPsi_g) = \N (\bx_{i};\bT \bsbeta_g,\sigma_g^2\Identity_m), 
\end{eqnarray}}where $\bsbeta_{g}$ is the coefficient  vector of the polynomial or spline regression model representing class $g$, $\sigma_{g}^2$ the associated noise variance and $\bT$ is the matrix of design which depends on the adopted model (e.g., Vandermonde matrix for polynomial regression). 
%
%
%
A similar approach that fits a specific regression model governed by a hidden logistic process to each of the $G$ homogeneous classes of curves presenting regime changes has been presented in \cite{chamroukhi_et_al_neurocomputing2010}.
However, all these approaches, as they involve a single model for each class, are only suitable for homogeneous classes of curves. For complex-shaped classes, when one or more classes are dispersed, the hypothesis of a single model description for the whole class of curves becomes restrictive.
This problem can be handled, by analogy to mixture discriminant analysis for vectorial data \cite{hastieANDtibshiraniMDA}, by adopting a mixture model formulation in the functional space. This functional mixture can for example be a polynomial regression mixture 
 or a spline regression mixture 
 \cite{chamroukhi_PhD_2010, Gui_FMDA}. 
  This leads to Functional Mixture Discriminant Analysis (FMDA).
%

\subsection{Functional Mixture Discriminant Analysis with  polynomial regression and spline regression mixtures}
\label{FMDA from the state of the art}
 
A first idea on FMDA, based on B-spline regression mixtures, was proposed in \cite{Gui_FMDA}. Each class $g$ of functions is modeled as a mixture of $K_g$ sub-classes, each sub-class $k$ ($k=1,\ldots,K_g$), is a noisy  B-spline function (and can also be a polynomial or spline) with parameters $\bsPsi_{gk}$. 
The model is therefore defined by:
{\small \begin{equation}
p(\bx_i|y_i = g, \bt; \bsPsi_g) 
= \sum_{k=1}^{K_g} \alpha_{gk} \N (\bx_{i};\bT \bsbeta_{gk},\sigma_{gk}^2\Identity_m), 
\label{eq: class mixture density for classic FMDA}
\end{equation}}where 
the $\alpha_{gk}$'s are the corresponding  non-negative mixing proportions that sum to 1,
 $z_i$ is a hidden discrete variable in $\{1,\ldots,K\}$ representing the labels of the sub-classes for each class.  
The parameters $\bsPsi_g = (\alpha_{g1},\ldots,\alpha_{gK_g}, \bsPsi_{g1},\ldots,\bsPsi_{gK_g})$ of this functional mixture density 
can be estimated by maximum likelihood using the EM algorithm \cite{dlr, Gui_FMDA}.  
%
%
However, using polynomial or spline regression for class representation, as studied in \cite{chamroukhi_PhD_2010, chamroukhi_et_al_neurocomputing2010} is more adapted for curves presenting smooth regime changes and the knots have to be fixed in advance. 
When the regime changes are abrupt, capturing the regime transition points needs to relax the regularity constraints on splines which leads to piecewise regression for which the knots can be optimized using a dynamic programming procedure. %
On the other hand, the regression model with a hidden logistic process (RHLP) presented in \cite{chamroukhi_et_al_neurocomputing2010} and used to model an homogeneous set of curves with regime changes, is flexible and explicitly integrates the smooth and/or abrupt regime changes via a logistic process. As pointed in \cite{chamroukhi_et_al_neurocomputing2010}, this approach has limitations in the case of complex-shaped classes of curves since each class is only approximated by a single RHLP model.
In this paper, we  extend the discrimination approach in \cite{chamroukhi_et_al_neurocomputing2010} to a functional mixture discriminant analysis framework, where each component density model for each homogeneous sub-class is assumed to be an RHLP model. 
We may therefore 
overcome both the limitation of FLDA and FQDA which are based only a single model for each class, by modeling each complex-shaped class of curves using a mixture of RHLP models (MixRHLP). Furthermore,  thanks to the flexibility to the RHLP model for each sub-class, we will be able to automatically and flexibly approximate the underlying regimes. 
%

 
\section{Proposed Functional Mixture Discriminant Analysis with hidden process regression mixture}
\label{FMDA from the state of the art}
%
 Let us suppose that each sub-class $k$ $(k=1,\ldots,K_g)$ of class $g$ itself is governed by $R_{gk}$ unknown regimes. We let therefore denote by $h_{gkj} \in \{1,\ldots,R_{gk}\}$ the discrete 	 variable representing the label of regimes for sub-class $k$ within class $g$.

%
%
\subsection{A mixture of RHLP models for each class of curves} 
In the proposed FMDA, we model each class of curves by a specific mixture of regression models with hidden logistic processes (abbreviated as MixRHLP) as in \cite{chamroukhi_PhD_2010,chamroukhi_adac_2011}. 
According to the MixRHLP model, 
 each class of curves $g$ is assumed to be composed of $K_g$ homogeneous sub-groups with prior probabilities 
 $\alpha_{g1},\ldots,\alpha_{gK_g}$. Each of the $K_g$ sub-groups is governed by $R_{gk}$ polynomial regimes and is modeled by a regression model with hidden logistic process (RHLP). 
The RHLP model \cite{chamroukhi_et_al_neurocomputing2010} assumes that each sub-class (or cluster) $k$ of class $g$ is  generated by $K_g$  polynomial regression models governed by a hidden logistic process $\bh_{gk}=(h_{gk1},\ldots,h_{gkm})$ that allows for  switching from one regime to another among $R_g$ polynomial regimes over time.   
%
%
Thus, the resulting conditional distribution of a curve $\bx_i$ issued from class $g$ is given by the following mixture:
{\small \begin{eqnarray}
p(\bx_i|y_i=g, \bt;\bsPsi_g)  
&=& \sum_{k=1}^{K_g} \alpha_{gk} \prod_{j=1}^m \sum_{r=1}^{R_{gk}}\pi_{gkr}(t_j;\bw_{gk})\mathcal{N}\big(x_{ij};\bsbeta_{gkr}^T \bt_{j},\sigma_{gkr}^{2} \big)
\label{eq: MixRHLP}
\end{eqnarray}}where $\bsPsi_g =(\alpha_{g1},\ldots,\alpha_{gK_g},\bstheta_{g1},\ldots,\bstheta_{gK_g})$ is the parameter vector of class $g$,  
 $\bstheta_{gk} = (\bw_{gk},\bsbeta_{gk1},\ldots,\bsbeta_{gkR_{gk}},\sigma_{gk1}^{2},\ldots,\sigma_{gkR_{gk}}^{2})$ being the parameters of each of its 
%
RHLP component density $\prod_{j=1}^m \sum_{r=1}^{R_{gk}}\pi_{gkr}(t_j;\bw_{gk})\mathcal{N}\big(x_{ij};\bsbeta_{gkr}^T \bt_{j},\sigma_{gkr}^{2} \big)$ 
where  
$\pi_{gkr}(t_j;\bw_{gk})$ represents the  probability of regime $r$ within sub-class $k$ of class $g$ and is modeled by a logistic distribution, that is \linebreak  
{\small $\pi_{gkr}(t_j;\bw_{gk})= p(h_{gkj}=r|t_j;\bw_{gk})=\frac{\exp{(w_{gkr0} + w_{gk1}t_j)}}{\sum_{\ell=1}^{R_{gk}}\exp{(w_{g\ell r 0} + w_{g \ell r 1} t_j)}},$} with parameters 
 $\bw_{gk} = (\bsw_{gk1},\ldots,\bsw_{gkR_{gk}})$ 
where  $\bsw_{gkr}=\{w_{gkr0},w_{gkr1}\}$. 
The relevance of the logistic process in terms of flexibility of transition has been well detailed in \cite{chamroukhi_et_al_neurocomputing2010}.
Notice that the key difference between FMDA with hidden process regression and FMDA of \cite{Gui_FMDA} is that the proposed approach uses a generative hidden process regression model (RHLP) for each sub-class rather than a spline; the RHLP being itself based on a mixture formulation. 
Thus, The proposed approach is more adapted for capturing the regime changes within curves. 
%
%
%
%
\vspace*{-.2cm}
\subsection{Maximum likelihood  estimation via the EM algorithm}
\label{sec: parameter estimation by EM mixture functional rhlp} 

The parameter vector $\bsPsi_g$ of the mixture density of class $g$ in Eq. \ref{eq: MixRHLP} is estimated by maximizing the observed-data log-likelihood. Given an independent training set of labeled curves, the log-likelihood of $\bsPsi_g$ is written as:
\vspace*{-.1cm}
{\small \begin{equation}
\cL(\bsPsi_g)
=\sum_{i|y_i=g} \log \sum_{k=1}^{K_g} \alpha_{gk} \prod_{j=1}^m \sum_{r=1}^{R_{gk}}\pi_{gkr}(t_j;\bw_{gk})\mathcal{N}\big(x_{ij};\bsbeta_{gkr}^T \bt_{j},\sigma_{gkr}^{2} \big).
\label{eq: loglik mixture fucntion rhlp}
\vspace*{-.1cm}
\end{equation}}The maximization of this log-likelihood 
is performed 
iteratively by a dedicated EM algorithm \cite{dlr}. 
The EM algorithm starts with an initial parameter $\bsPsi_g^{(0)}$ and alternates between the two following steps until convergence:
\vspace*{-.3cm}
\paragraph{{\bf E-step:}}
\label{par: E-step mixture of rhlp and EM}
Compute the expected complete-data log-likelihood given the observations $(\{\bx_i|y_i=g\},\bt)$ and the current parameter estimation  $\bsPsi_g^{(q)}$: 
{\small \begin{eqnarray}
Q(\bsPsi_g,\bsPsi_g^{(q)})& \!\!\! = \!\!\!  & \E\left[\cL_c(\bsPsi_g;\{\bx_i|y_i=g\},\bz,\{\bh_{gk}\},\bt)|\{\bx_i|y_i=g\},\bt;\bsPsi_g^{(q)}\right].
\label{eq: Q-function for the MixRHLP}
\end{eqnarray}}
This step simply requires the calculation of the posterior sub-class probabilities: 
$\gamma_{igk}^{(q)} = \frac{\alpha_{gk}^{(q)}\prod_{j=1}^m\sum_{r=1}^{R_{gk}}\pi_{gkr}(t_j;\bw_{gk}^{(q)})\mathcal{N}\big(x_{ij};\bsbeta^{T(q)}_{gkr}\bt_{j},\sigma^{2(q)}_{gkr}\big)}
{ \sum_{l=1}^K \alpha_{gl}^{(q)}\prod_{j=1}^m\sum_{r=1}^{R_{gl}} \pi_{glr}(t_j;\bw_{gl}^{(q)}) \mathcal{N}(x_{ij};\bsbeta^{(q)T}_{glr}\bt_{j},\sigma^{2(q)}_{glr})}$
and the posterior regime probabilities for each sub-class: 
$\tau^{(q)}_{ijgkr} = \frac{\pi_{gkr}(t_j;\bw_{gk}^{(q)})\mathcal{N}(x_{ij};\bsbeta^{T(q)}_{gkr}\bt_{j},\sigma^{2(q)}_{gkr})}
{\sum_{\ell=1}^{R_{gk}}\pi_{g\ell r}(t_j;\bw_{g\ell}^{(q)})\mathcal{N}(x_{ij};\bsbeta^{T(q)}_{g\ell r}\bt_{j},\sigma^{2(q)}_{g \ell r})}\cdot$
\vspace*{-.2cm}
\paragraph{{\bf M-step:}}
\label{par: M-step mixture of rhlp and EM}
Update the value of the parameter $\bsPsi_g$ by maximizing the function $Q(\bsPsi_g,\bsPsi_g^{(q)})$ with respect to $\bsPsi_g$, that is: $\bsPsi_g^{(q+1)} = \arg \max_{\bsPsi_g} Q(\bsPsi_g,\bsPsi_g^{(q)})$. 
It can be shown that the maximization of the $Q$-function  can be performed by separate  maximizations w.r.t the mixing proportions $(\alpha_{g1},\ldots,\alpha_{gK_g})$ 
and w.r.t the regression parameters $\{\bsbeta_{gkr},\sigma^2_{gkr}\}$ and the hidden logistic process parameters $ \{\bw_{gk}\}$. The mixing proportions updates are given by $\alpha_{gk}^{(q+1)} = \frac{1}{n_g}\sum_{i|y_i=g} \gamma_{igk}^{(q)}$. 
The maximization w.r.t the regression parameters 
 consists in performing separate analytic solutions of weighted least-squares problems where the weights are the product of the posterior probability $\gamma^{(q)}_{igk}$ of sub-class $k$ and the posterior probability $\tau^{(q)}_{ijgkr}$ of regime $r$ within sub-class $k$. The updates are: 
\vspace{-.3cm}
{\small \begin{eqnarray}
\!\!\!\!\!\! \bsbeta^{(q+1)}_{gkr} &\!\!\!\! = \!\!\!\! & \Big[\sum_{i|y_i=g} \sum_{j=1}^{m} \gamma_{igk}^{(q)} \tau^{(q)}_{ijgkr}\bt_j \bt_j^T\Big]^{-1}\sum_{i|y_i=g} \sum_{j=1}^{m} \gamma_{igk}^{(q)} \tau^{(q)}_{ijgkr}x_{ij} \bt_j  \nonumber\\ 
\!\!\!\!\!\! \sigma_{gkr}^{2(q+1)} &\!\!\!\! =\!\!\!\! & \frac{1}{ \sum_{i|y_i=g}\sum_{j=1}^m \gamma_{igkr}^{(q)}\tau_{ijgkr}^{(q)} }\sum_{i|y_i=g}\sum_{j=1}^{m} \gamma_{igkr}^{(q)}\tau^{(q)}_{ijgkr} (x_{ij}-{\bsbeta}^{T(q+1)}_{gkr} \bt_j)^2 \cdot 
\label{eq: EM estimate of variance sigma^2_rk for polynom k of the sub-class r for the MixRHLP}
\end{eqnarray}}Finally, the maximization 
 w.r.t the logistic processes parameters $\{\bw_{gk}\}$ 
 consists in solving multinomial logistic regression problems weighted by $\gamma_{igk}^{(q)}\tau^{(q)}_{ijgkr}$ which we solve with a multi-class IRLS algorithm (e.g., see  \cite{chamroukhi_PhD_2010}). 

\section{Experiments on simulated curves} 
\label{sec: Experiments on simulated curves}

\begin{figure}[!h]
 \centering
 \includegraphics[width=2.35cm]{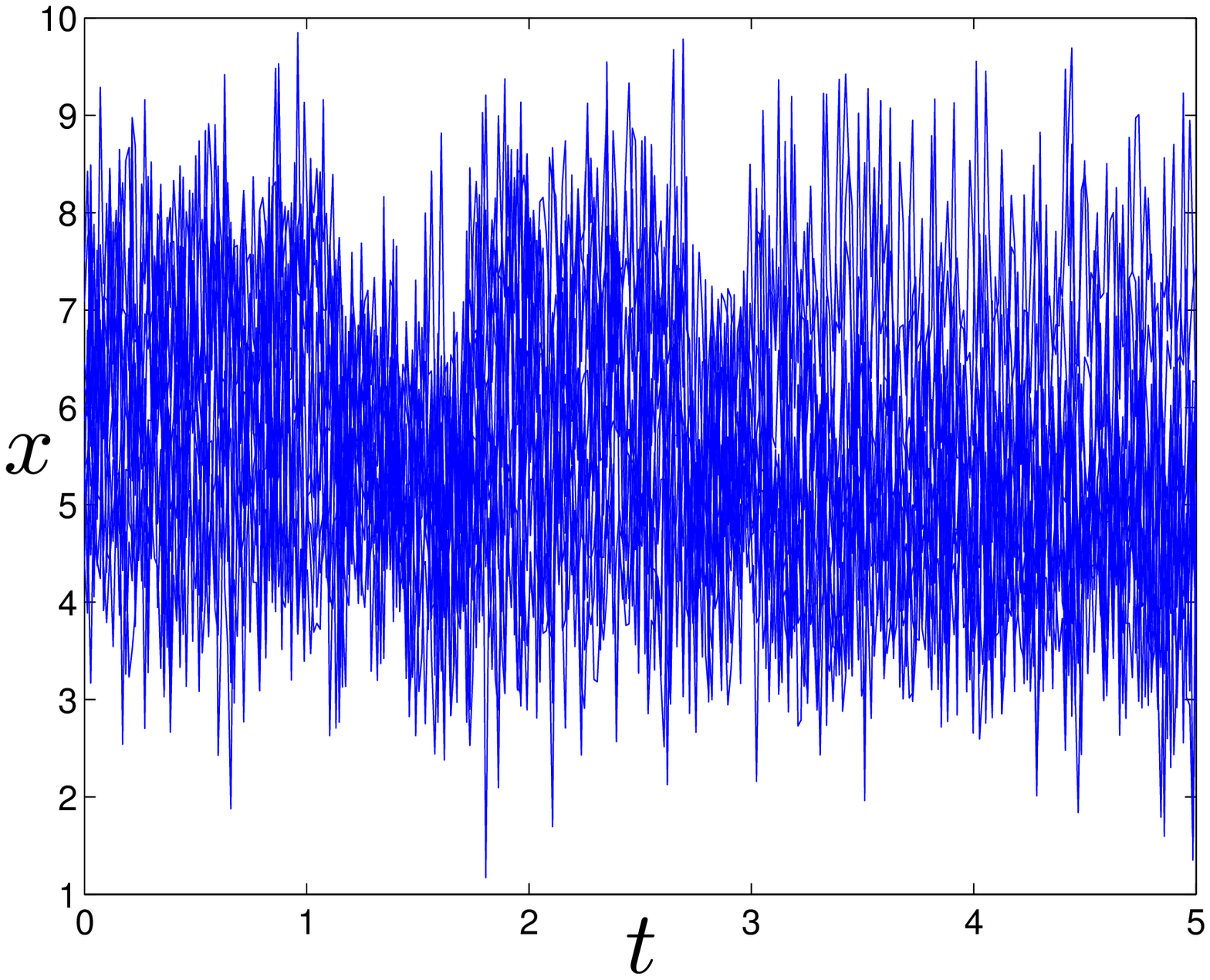}  
 \includegraphics[width=2.35cm]{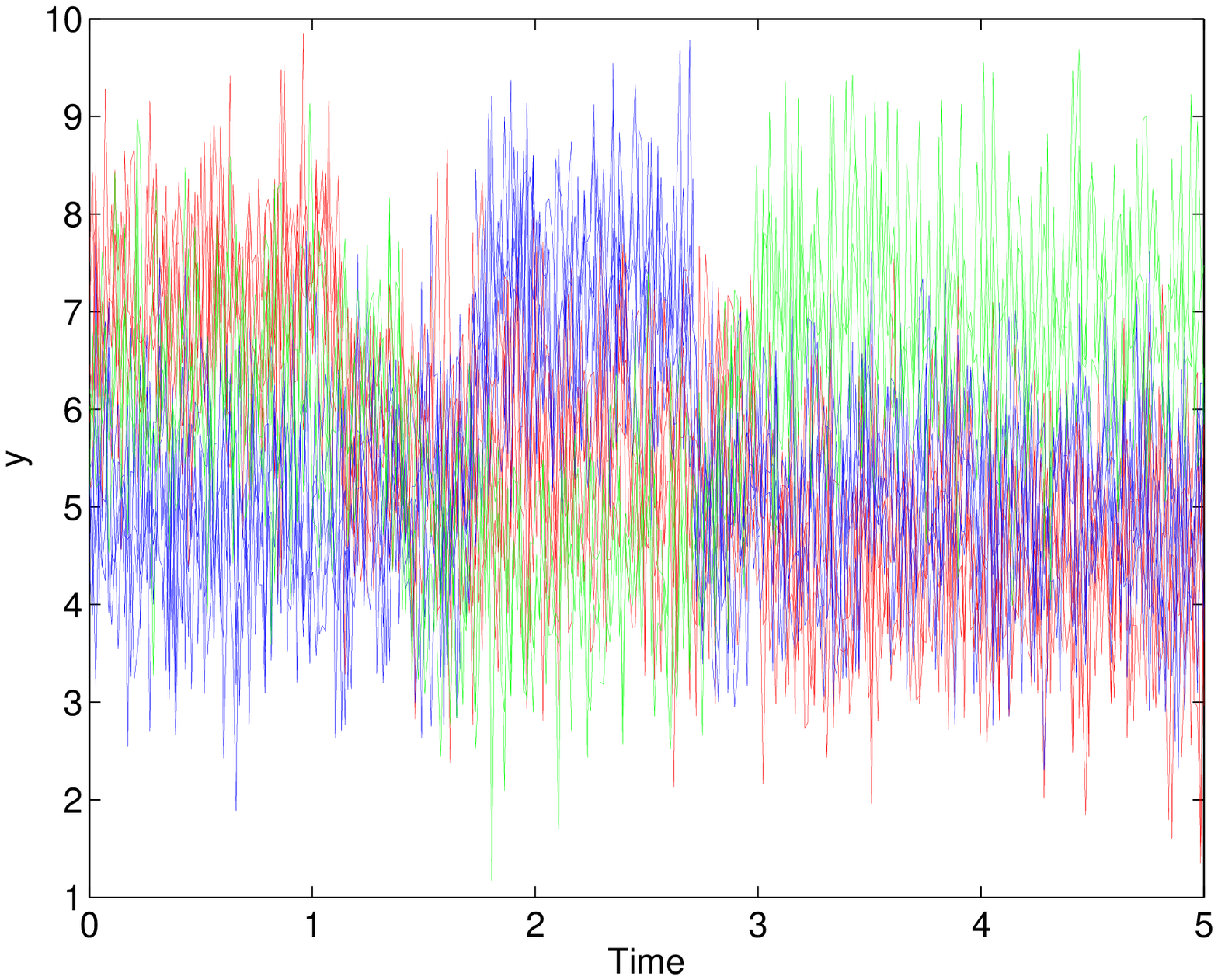} 
 \includegraphics[width=2.3cm]{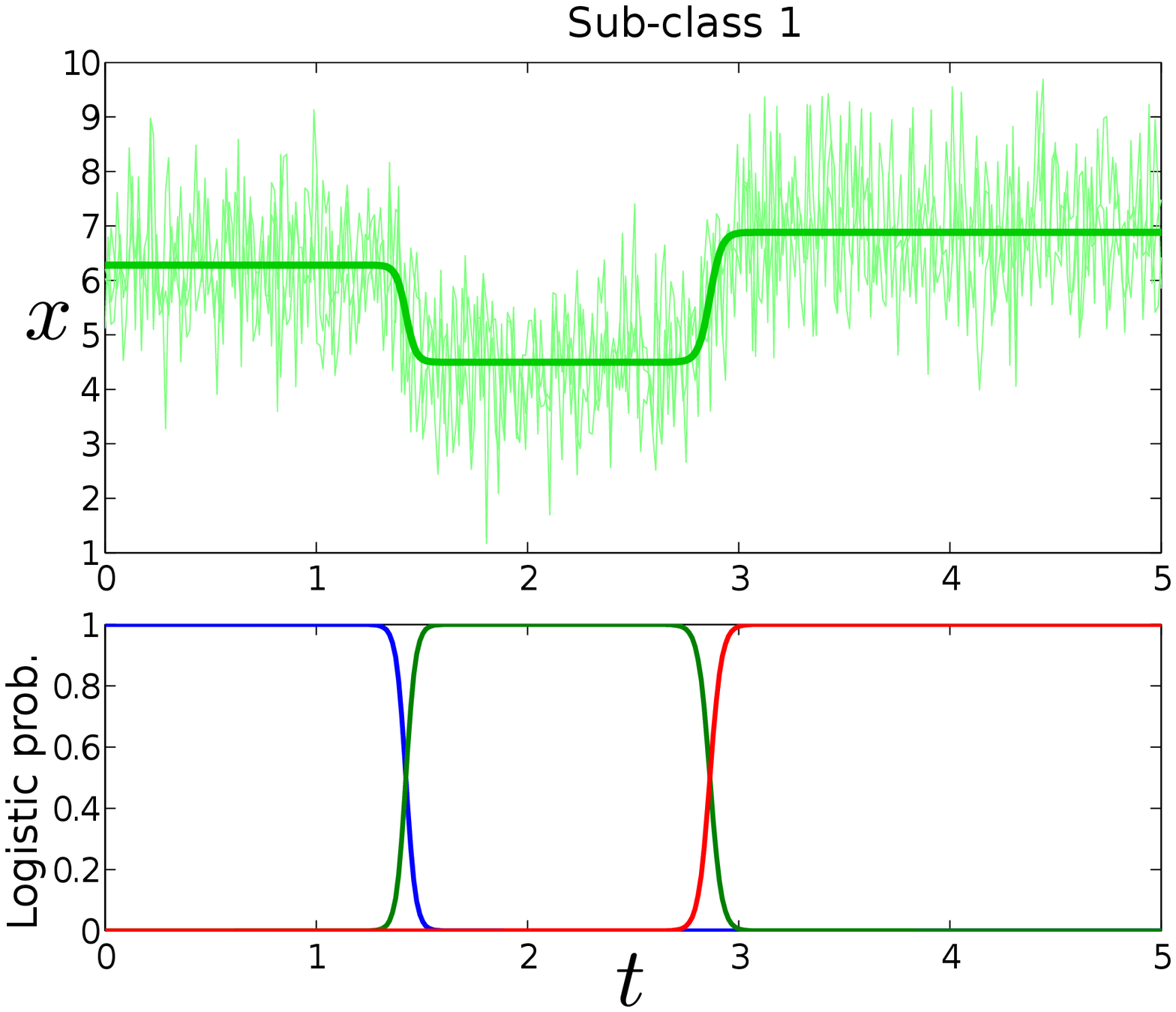} 
 \includegraphics[width=2.3cm]{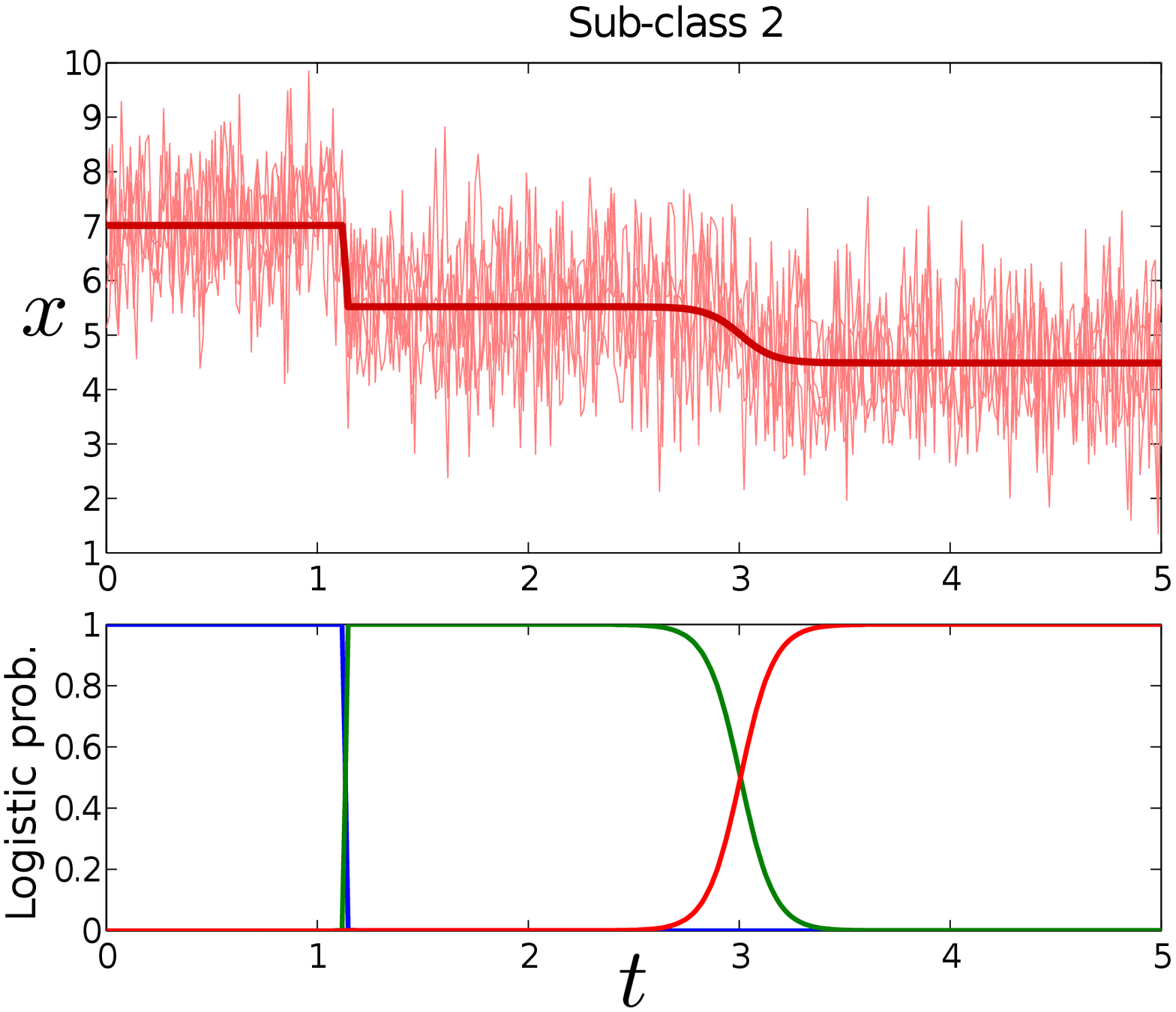}
 \includegraphics[width=2.3cm]{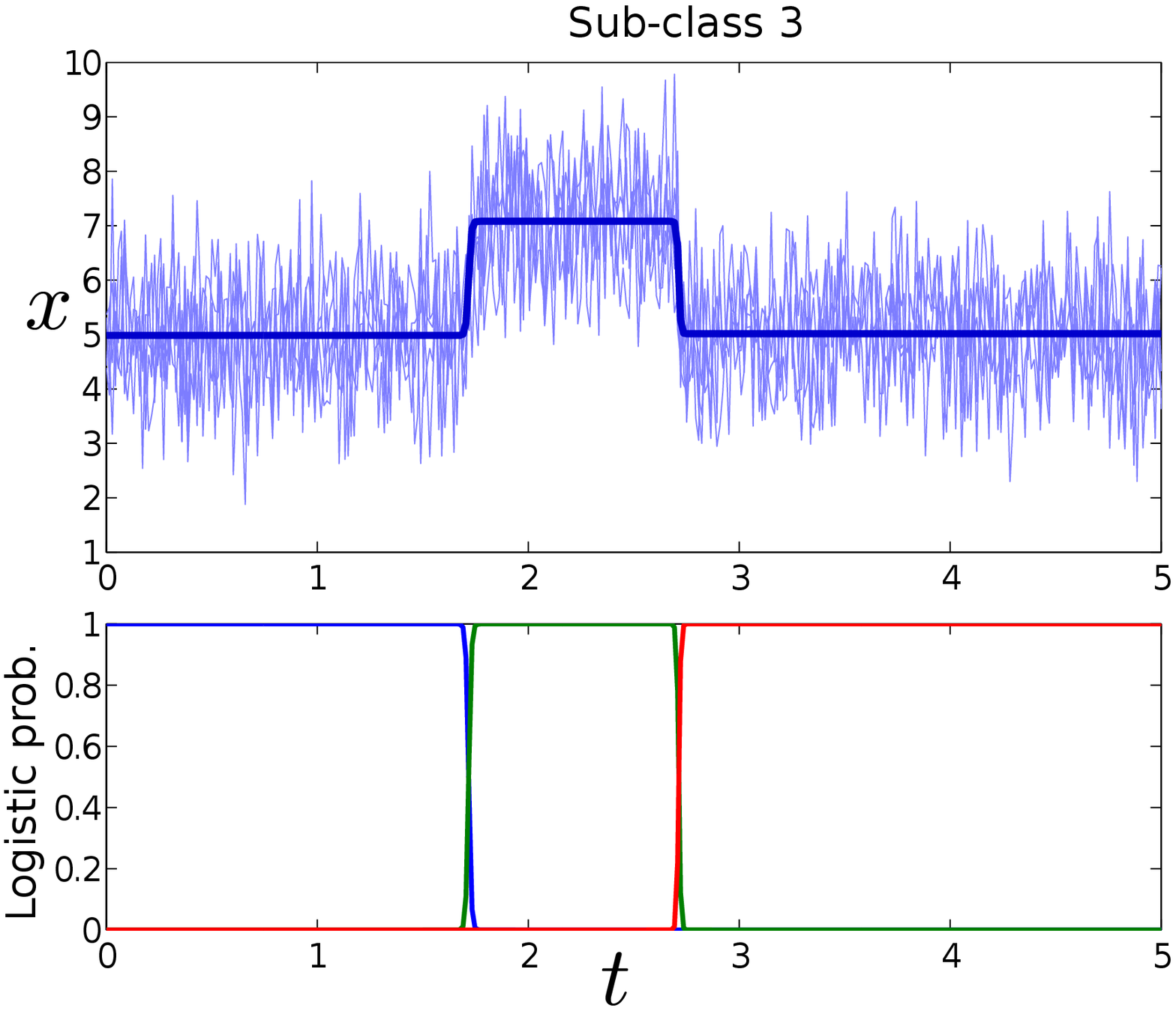}\\
  \vspace{-.2cm}
{\small \caption{\label{fig: simulated curves and results}
The complex-shaped class of curves (left) and the  three sub-classes 
  obtained by the functional mixture model with hidden process regression.}}
\end{figure}

 We perform comparisons between the proposed approach and alternative functional discriminant analysis approaches 
and functional mixture discriminant analysis approaches. 
The used criterion is the curve misclassification error rate computed by a $5$-fold cross-validation procedure.
%
We consider simulated curves issued from two classes of piecewise noisy curves. The first class is composed of three sub-classes (see Fig \ref{fig: simulated curves and results}), while the second one is a homogeneous class. Each curve is composed of $m=200$ points and consists of three piecewise constant regimes.
Table \ref{table: classification results for simulated curves} shows the misclassification error rates 
 obtained with the proposed FMDA approach and the alternative approaches. 
 As expected, it can be seen that the FMDA approaches provide better results compared to FLDA approaches. This is due to the fact that the class shape  is  complex (see Fig \ref{fig: simulated curves and results}) to be approximated by a single model as in FLDA. It can also be observed that 
the proposed functional mixture discriminant approach based on hidden logistic process regression outperforms the alternative FMDA based on polynomial or spline regression. This performance is attributed to the flexibility of the logistic process that it is well adapted for the regime changes.
%
 

%
%
  \vspace{-.8 cm}
\begin{table}[!h]
\centering
{\tiny \begin{tabular}{|c|l|c|} 
\hline
\multicolumn{2}{|c|}{Discrimination Approach} &  Misclassification error rate (\%)\\ 
\hline
\hline
	& Polynomial regression   & 21 \ $\pm$(5.5) \\ 
FLDA& Spline regression  & 19.3 $\pm$(6.5)\\ 
	& Hidden process regression (RHLP) & 18.5 $\pm$(4) \\ 
\hline
	& Polynomial Regression Mixture & 11$\pm$(5.94) \\ 
FMDA & Spline Regression Mixture     & 9.5$\pm$(4.61) \\ 
	& {\bf Proposed FMDA with MixRHLP }       & 5.3$\pm$(2.4) \\ 
	\hline
\end{tabular}}
  \vspace{-.1cm}
\caption{\label{table: classification results for simulated curves}
Obtained discrimination results for the simulated curves.}
\end{table}

\section{Conclusion}
\label{sec: conclusion}
We presented a new approach for functional data classification by using a mixture model-based functional mixture discriminant analysis 
which is based on a hidden process regression model. 
Each class parameters are estimated  by a dedicated EM algorithm.
%
%
The experimental results on simulated data demonstrated the benefit of the proposed approach to addressing the problem of modeling complex-shaped classes of curves as compared to existing alternative functional discriminant methods. 
Future work will concern experiments on real data including gene expression curves and 2-d 
functions from handwritten digits. We will as well investigate using Bayesian learning 
to better control the model complexity. 


\vspace*{-.2cm}
\begin{footnotesize}

\end{footnotesize}


\end{document}